# Optical excitation and detection of high-frequency Sezawa modes in Si/SiO$_2$ system decorated with Ni$_{80}$Fe$_{20}$ nanodot arrays

P. Graczyk,[1][*] B. Rana,[2,3][*][a] A. Trzaskowska,[3] B. K. Mahato,[2,4] J. W. Kłos,[3] M. Krawczyk,[3] A. Barman[2][b]

[1] Institute of Molecular Physics, Polish Academy of Sciences, M. Smoluchowskiego 17, 60-179 Poznan, Poland
[2] Department of Condensed Matter and Materials Physics, S. N. Bose National Centre for Basic Sciences, Block JD, Sector III, Salt Lake, Kolkata 700 106, India
[3] Institute of Spintronics and Quantum Information, Faculty of Physics and Astronomy, Adam Mickiewicz University, Poznan, Uniwersytetu Poznańskiego 2, 61-614 Poznań, Poland
[4] National Institute of Education, Nanyang Technological University, 1 Nanyang Walk, Singapore 637616, Singapore

[*] These Authors contributed equally to the research.
[a] Corresponding Author: bivran@amu.edu.pl
[b] Corresponding Author: abarman@bose.res.in

## Abstract

Surface acoustic waves have emerged as one of the potential candidates for the development of next-generation wave-based information and computing technologies. For practical devices, it is essential to develop the excitation techniques for different types of surface acoustic waves, especially at higher microwave frequencies, and to tailor their frequency versus wave vector characteristics. We show that this can be done by using ultrashort laser pulses incident on the surface of a multilayer decorated with a periodic array of metallic nanodots. Specifically, we study surface acoustic waves in the dielectric substrate Si/SiO$_2$ decorated with a square lattice of thin Ni$_{80}$Fe$_{20}$ (Py) dots. Using a femtosecond laser-based optical pump-probe measurement, we detect a number of high-frequency phononic modes. By performing finite element simulations, we identify them as Sezawa modes from the second and third Brillouin zone in addition to the modes confined within the Py dots. The frequency of the Sezawa modes strongly depends on the period of the Py dots and varies in the range between 5 to 15 GHz. Both types of waves cover the same frequency range for Py dots with period less than 400 nm, providing a promising system for magnetoelastic studies.

# 1. Introduction

Acoustic waves (phonons) are excellent carriers for Gigahertz (GHz)-frequency signals because of their linear dispersion relations, which limits the spatial dispersion of wave packets, and their low damping, which ensures long-range propagation. The phononic devices have been extensively studied and used for a long time [1], [2], [3], but only recent developments in nanofabrication techniques allow the integration of phononic systems with micrometer-sized electronic circuits operating at GHz frequencies [4], [5], [6], [7]. In these applications, planar geometries with surface acoustic waves (SAWs) are widely used.

In multilayered materials, one can distinguish Rayleigh modes, Sezawa modes (also known as a higher-order Rayleigh modes) and Love modes among the SAWs [8]. Rayleigh and Sezawa modes are polarized in the sagittal plane, i.e., the plane containing the wavevector and the vector normal to the surface, while Love modes are the horizontal shear waves. Sezawa and Love modes appear in the substrate-layer systems where the transverse acoustic wave velocity of the top layer is lower than that in the substrate (or sub-layers), and the subsequent modes are characterized by the increasing number of nodal lines along the depth of the system. Sezawa modes have gained increased interest due to their higher frequency than Rayleigh mode and more efficient generation, detection and sensitivity to the external environment than Love modes. This interest is reflected by the use of the Sezawa modes in many applications, e.g. filters, phase shifters, resonators, microfluidic systems and various types of sensors [9], [10].

SAWs are usually generated by the electromechanical coupling, using an interdigital transducer antenna placed on top of a piezoelectric layer. For the high-frequency applications, it is necessary to use advanced lithography techniques to achieve suitably narrow antennas, whose width matches the low SAW wavelengths. However, the resistivity of the nanometric transmission lines increases as a result of increased contribution of the electron scattering from surfaces and grain boundaries [11], [12]. Therefore, high-quality single-crystal wires are required [13]. An alternative strategy to enhance the amplitude of the higher-order modes supported by the wide antenna is to use a phononic crystal, i.e., a periodically repeated pattern in nanoscale [14], as a grating coupler. It enhances the amplitude of the short-wavelength waves compared to the bare antenna [15]. In 2007 Robillard et al., [16], [17], [18] showed experimentally that it is possible to generate and detect multiple Rayleigh waves by the spatially quasi-uniform optical pulse in the two-dimensional phononic crystal made of metallic nanodots. The wavenumbers of the excited waves were defined by phase matching condition for SAW wavenumber, $k = 2\pi n/a$, where $a$ is the lattice constant and $n$ is an integer. That is, the excited Rayleigh waves are the waves that lay at the center of the Brillouin zone (BZ) of the phononic crystal, i.e., inside the sound cone. This implies that these are leaky SAWs, coupled to the bulk modes.

In this paper, we propose a system for a grating-assisted excitation and detection of the Sezawa modes. For this purpose, two criteria have to be considered: (1) the system must be a so-called slow-on-fast system, i.e., the bulk shear velocity $v$ of the layer must be lower than that of the substrate, to allow the existence of the Sezawa modes, and (2) the cutoff wavenumber [19] of (at least) the first Sezawa mode must be lower than the highest wavenumber accessible in the experiment, which is defined similarly to Refs [15-16] by the lattice constant of the periodic pattern. Specifically, we chose an Si/SiO$_2$ system, which is characterized by a rather high velocity contrast: $\frac{v_{\text{SiO2}}}{v_{\text{Si}}} = 0.64$ thus satisfying criterion (1). Next, the cutoff wavenumber of the first Sezawa mode can be estimated from the formula [19], [20]:

$$k_c = \frac{2\pi v_{\text{SiO2}}}{4h\sqrt{v_{\text{Si}}^2 - v_{\text{SiO2}}^2}} = 6.75 \text{ rad µm}^{-1},$$

where $h = 300$ nm is the thickness of the SiO$_2$ layer. Then, for the periodic pattern of the lattice constant $a < a_c = \frac{2\pi}{k_c} = 930$ nm, the first Sezawa mode should be accessible. The second Sezawa mode is expected to be accessible at three times smaller lattice constant. We used metallic permalloy (Ni$_{80}$Fe$_{20}$, Py) nanodots to create the periodic pattern, because this material has even lower shear velocity, thus inducing more load on the substrate and lowering cutoff wavenumber for Sezawa modes. In addition, it is a ferromagnetic metal that offer suitable condition for spin wave excitations [21], which may allow future exploration of magnetoelastic effects in this system [22], [23], [24], [25].

Therefore, here we experimentally study the frequency spectra of SAWs in structures composed of a homogeneous Si/SiO$_2$ substrate decorated with nanometer-sized square Py dots arranged in a square lattice, and thus forming a phononic crystal. We fabricated a series of samples with fixed dot sizes and lattice constant $a$ ranging from 275 nm to 600 nm ($a < a_c$), covering accessible wavenumbers from 10 rad µm$^{-1}$ to 23 rad µm$^{-1}$. We used the optical pump-probe technique, commonly used for the measurements of elastic and spin-wave excitations [26], [27], [28], which is based on the time-domain measurements of the Brillouin scattering of light on elementary excitation of the solid. We show that decorating the substrate with such phononic crystals, allows us to optically excite and detect in reflection the Rayleigh mode from the first Brillouin zone, two Sezawa modes from the edge of the second Brillouin zone, and additionally, two non-dispersive modes localized in the Py dots. These interpretations of the experimental results are based on Finite Element Method (FEM) simulations, which are also used to calculate the full band structure of the SAWs in the considered systems.

## 2. Sample structure and methods

**Sample fabrication:**

10 µm × 10 µm square lattices of Py elements with a size (*W*) of 200 nm × 200 nm, thickness (*t*) of 50 nm and interelement (edge-to-edge) separation (*S*) varying from 75 to 400 nm (i.e. the lattice constant $a = S + W$ varying from 275 to 600 nm) were prepared by a combination of electron beam lithography and electron beam evaporation. At first, bilayer PMMA/MMA (polymethyl methacrylate/methyl methacrylate) resist was coated on Si[100]/SiO$_2$(300 nm) substrate by a spin coater. The patterns of dot arrays were then prepared on the coated resists by electron beam lithography with a dose current of 100 pA and a dose time of 0.8 µs. Finally, 50-nm-thick Py film was deposited on the resist pattern by electron beam evaporation at a base pressure of about $2 \times 10^{-8}$ Torr. A 10-nm-thick SiO$_2$ capping layer was also deposited on the top of Py to protect the dots from degradation when exposed to air during optical pump-probe measurements. This was followed by the lifting-off of the sacrificial material and oxygen plasma cleaning of the residual resists that remained even after the lift-off process. Figure 1 represents scanning electron micrographs of the studied samples.

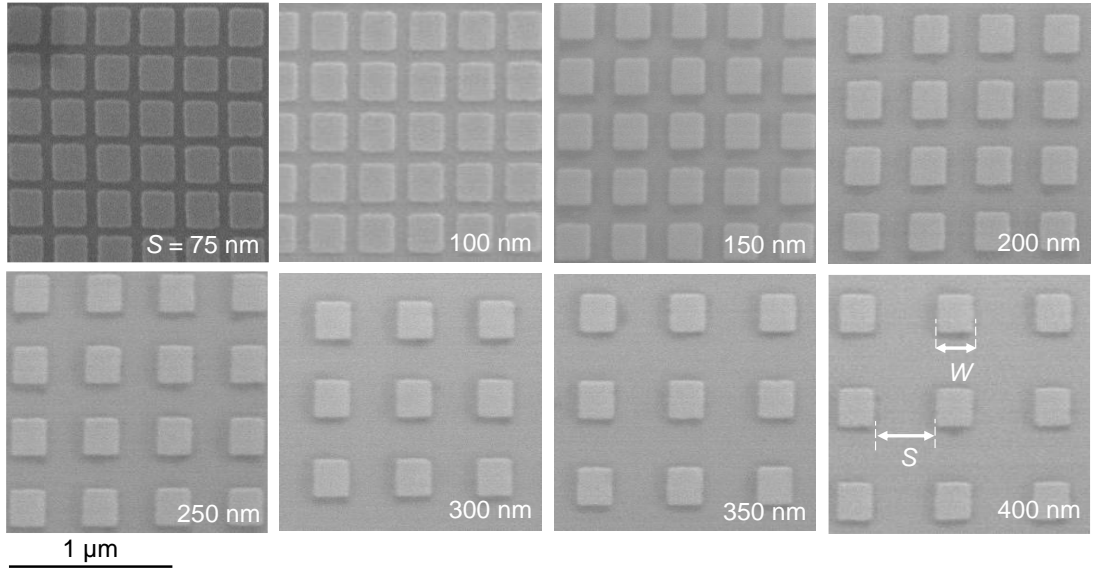

**Fig. 1:** The scanning electron micrographs of the arrays of Py elements with size ($W$) of 200 nm × 200 nm, thickness ($t$) of 50 nm and interelement (edge-to-edge) separation ($S$) varying from 75 to 400 nm.

**Measurement procedure:**

The femtosecond pulsed laser beam induced SAWs were measured by using a home-built time-resolved magneto-optical Kerr effect microscope based upon a two-color collinear pump-probe setup (see, Fig. 2). The second harmonic ($\lambda$ = 400 nm, pulse width ≈ 100 fs) of a Ti-sapphire laser beam (Tsunami, Spectra Physics, pulse width ≈ 80 fs) was used to excite SAWs, while the time delayed fundamental ($\lambda$ = 800 nm) laser beam was used to probe SAWs by measuring the total reflectivity signals by means of an optical bridge detector, i.e., balanced photodetector, which enabled us to isolate the total reflectivity signals from magnetic Kerr signals. The variation of the reflectivity in time is induced because of the interference of the elastically scattered light with the inelastically scattered light via Brillouin scattering, i.e., mainly surface ripple mechanism [29]. The interference produces small intensity variations at the photodetector. This experimental technique is known in the literature as picosecond acoustic interferometry or time-domain Brillouin scattering (TD-BS) [30].

When femtosecond laser beam interacts with the Si/SiO$_2$ film decorated with Py nanodots, the exposed area experiences thermal expansion. This out of equilibrium disturbance induces elastic waves in the Py nanodots and Si/SiO$_2$ (i.e. at the Si/SiO$_2$ interface, because of high transparency of SiO$_2$). However, there is a difference between the way the laser beam induces elastic waves in (1) Py nanodots, and at (2) Si/SiO$_2$ interface. For case (1), because of small penetration depth, the laser beam is primarily confined within ~ 15 nm on top surface of Py nanodots. The heat generated initially on the upper part of Py nanodots is quickly distributed into the whole volume within several picoseconds due to high thermal conductivity of Py (or even faster due to the nonequilibrium diffusion and thermalization of photoexcited electrons in metals [31]). This causes homogeneous thermal expansion of Py nanodots. In this case the wavelengths of the elastic waves depend upon the periodicity of the nanodots.

For case (2), the laser beam induces thermal expansion at Si/SiO$_2$ interface and generates low frequency propagating Rayleigh waves. Here, the wavelengths of the excited Rayleigh waves are determined by the spot size of pump beam. The pump power used in these measurements is about 15 mJ cm$^{-2}$, while the probe power is much weaker and is about 2 mJ cm$^{-2}$. The pump beam was focused to a spot size of about 1 μm at the center of each array by a microscope objective with numerical aperture (N.A.) 0.65 and a closed loop piezoelectric scanning *x-y-z* stage. The probe beam was spatially overlapped with the pump beam after passing through the same microscope objective in a collinear geometry. Consequently, the pump spot was slightly defocused (spot size ≈ 1 μm) on the sample plane, which is also the focal plane of the probe spot. The pump beam was chopped at 2 kHz frequency, which was used as the reference for lock-in-amplifier detection of the reflectivity signal. The details of the experimental set up can also be found in Refs. [32], [33].

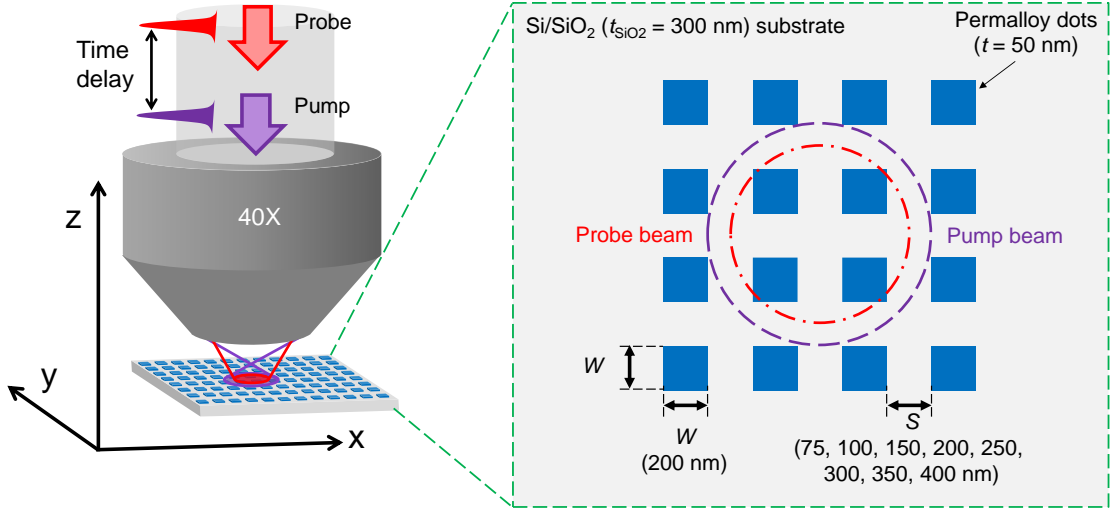

**Fig. 2.** Schematic diagram shows the geometry for all-optical measurement of femtosecond laser beam induced SAWs on a Si/SiO$_2$ substrate containing periodic arrays of 50-nm-thick and 200-nm-wide square Py dots with varying lattice constant. The collinear pump and probe beams are focused onto the sample with a spot size of about 1 μm and 800 nm, respectively, by using a 40x microscope objective with numerical aperture of 0.65. The inset shows the schematic diagram of a square array of square Py dots and focused spots of pump and probe beams onto it.

**Computations:**

The SAWs propagating in the structure were simulated by solving the elastic wave equation in the frequency domain by FEM using COMSOL Multiphysics [34], [35], [36], [37]. The elementary cell used in the FEM simulations is shown in Fig. 3. The silicon (Si) substrate was treated as a half-space ($z \leq 0$) with Py square dots on top. The 10-nm-thick SiO$_2$ capping layer was neglected in the simulations, as we verified that it does not affect the results but it significantly increases the computation time. The elastic constants and densities for each material used in the simulations are listed in Table 1. The Si substrate is a cubic crystal, while the other materials are assumed to be isotropic. The geometric parameters used in the simulations correspond to those of the experimental samples, but the simulations are performed for a larger number of Py-dot periods to obtain quasi-continuum changes of the SAW spectra with *a*. Thus, the size of the unit-cell was varied. The mesh size was 5 nm close to the surface and increased up to 10 $\mu$m with depth. The height of the elementary cell used in the simulations was correlated with the wavelength of the SAW. Dirichlet type of the boundary conditions were adopted at the bottom of the unit-cell and free boundary conditions were applied at the upper surface. Bloch-Floquet periodic boundary conditions were applied to the walls between the nearest unit cells in the *xz* and *yz* planes.

As an indicator of the SAW intensity $I(f_i, \mathbf{k}_i)$ measured in experiment we use the integral of the $z$ component of the displacement vector $u_z^{f_i,\mathbf{k}_i}$ for the selected mode $i$, at the selected frequency $f_i$ and wavevector $\mathbf{k}_i$ over the free surface of the Py dot $A$ [29], [38]:

$$I(f_i, \mathbf{k}_i) = \left| \int_A u_z^{f_i,\mathbf{k}_i} dA \right|^2. \qquad (1)$$

We used frequency-domain study with harmonic loads to reflect the experimental excitation of the system. Since our femtosecond pulse carries broad frequency spectrum up to the THz range, it is justified to assume that it is able to excite each frequency with the same efficiency and independently in the GHz range by thermal expansion. The loads $p$ are applied at the part of the material of the highest absorption of the electromagnetic wave, i.e., at the volume of the Py dots and at the Si/SiO$_2$ interface. The relative values of $p$ in those two regions (while the absolute values of $p$ are arbitrary in the linear problem) were deduced by attempting best fitting to the experimental spectra (Fig. 4) for $a = 275$ nm and $a = 600$ nm. The external stress applied in Py is $p = 100$ kN m$^{-2}$, while the boundary load at SiO$_2$/Si interface varies linearly from $p = 100$ kN m$^{-2}$ for $a = 600$ nm to $p = 10$ kN m$^{-2}$ for $a = 300$ nm. This reflects the fact that the coverage of the interior by Py dots increases with decreasing cell size. Damping is introduced to the whole system by setting the isotropic structure loss factor to $\eta_s = 0.01$. A perfectly matched layer condition is used in Si at the distance of 3 µm from the free surface. In these simulations we use $k_i = 0$ in Eq. (1) to obtain the SAW intensity. For simulation of the phononic dispersion relation and band structure we used eigenfrequency solver in COMSOL Multiphysics, with the same unit cell and parameters as in frequency domain, parametrized by the wavevector $\mathbf{k}_i$ along the irreducible path in the first Brillouin zone of the square lattice.

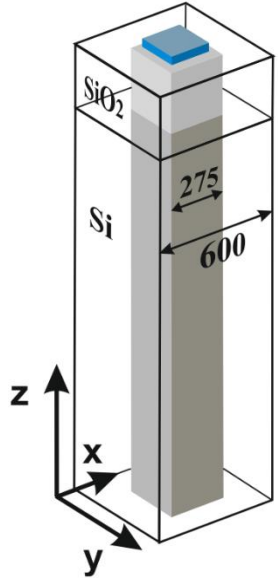

**Fig. 3**. Schematic view of the unit cell with periodic boundary conditions in the *xz* and *yz* planes used in the FEM simulations. The size of the unit cell was varied from 275 to 600 nm to reproduce the experimental results and in a wider range to deepen the analysis. The blue rectangular structure on the top surface of the unit cell represents a square Py dot.

**Table. 1.** Values of the physical parameters of the studied materials: $\rho$ – density (kg m$^{-3}$), $c_{ij}$ – elastic coefficients ($\times 10^{10}$ N m$^{-2}$), $E$ – Young modulus (GPa), and $\nu$ – Poisson ratio.

| | Silicon (Si) [2] | Silicon dioxide (SiO$_2$) [3] | Permalloy (Ni$_{80}$Fe$_{20}$) [4] |

| $\rho$ | 2331 | 2203 | 8690 |
|---|---|---|---|
| $c_{11}=c_{22}=c_{33}$ | 16.6 | -- | -- |
| $c_{44}=c_{55}=c_{66}$ | 8.0 | -- | -- |
| $c_{12}=c_{13}=c_{23}$ | 6.4 | -- | -- |
| $E$ | -- | 73.1 | 140 |
| $\nu$ | -- | 0.17 | 0.38 |

## 3. Results

In Figs. 4(a) and 4(b) we show the dispersion relation of the phonons calculated with the eigenfrequency solver in the limit systems, which are (a) the plain multilayer $Si/SiO_2$ without any pattern, i.e., the structure in the limit $q \equiv 2\pi/a \to 0$, and (b) $Si/SiO_2/Py$, i.e., in the limit $q \to \infty$, which can be related to the spectra of the patterned sample with very large and very small $a$, respectively. We see that the spectra are composed of the continuum bulk-band above the sound-line (hatched area), and some detached bands below this line which are SAWs. We distinguish Love modes from Rayleigh and Sezawa modes by calculating the SAW polarization $\xi$:

$$\xi = \frac{\int |u_y| dA}{\int u\, dA},$$

where $u_y$ is the y-component of the displacement and $u = \sqrt{u_x^2 + u_y^2 + u_z^2}$ is the total displacement and the integration is done on the free surface of the system. Therefore, $\xi$ has a value close to 1 for Love modes (blue color in Fig. 4), while it has a value close to 0 for Rayleigh modes (red color). As estimated in the introduction, the cutoff wavenumber of the first Sezawa mode is at $k \approx 6$ rad μm$^{-1}$ and the Py layer lowers the cutoff wavenumbers of Sezawa modes. Thus, from the results shown in Fig. 4, we expect to observe two or three Sezawa modes in addition to the Rayleigh mode in the range of $k$ from 10 rad μm$^{-1}$ to 20 rad μm$^{-1}$ in the experiment.

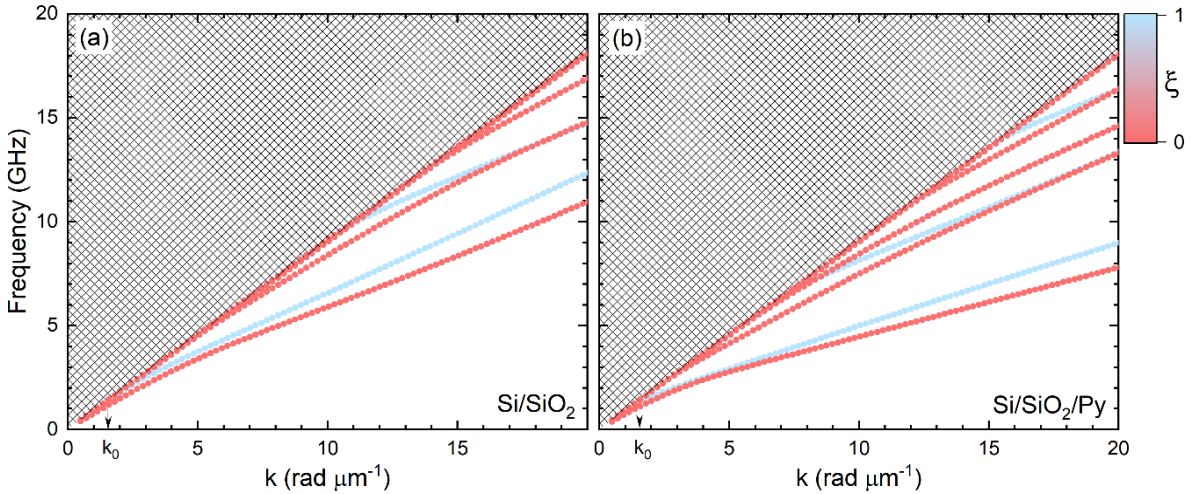

**Fig. 4.** Calculated dispersion relation of SAWs in the (a) plain $Si/SiO_2$(300 nm) and (b) $Si/SiO_2$(300 nm)/Py(50 nm) systems. The hatched pattern indicates the continuum of the bulk bands. The color map indicates the polarization $\xi$ of the surface wave, with blue corresponding to the Love modes, and red to the Rayleigh and Sezawa modes. The arrow marks the wavenumber selected by the experimental setup.

The results of measurements for the nanodot arrays with $a$ = 275, 300, 350, 400, 450, 500, 550 and 600 nm, are collected in Fig. 5. In the left panel the reflectivity spectra recorded as a function of time delay between pump and probe are presented up to a time delay of 2 ns. In the right panel their Fast Fourier Transformed (FFT) spectra are shown, with intensities normalized to their maximal values. It can be seen that the lowest frequency acoustic mode with maximum intensity is almost independent on the lattice constant, and varies between 0.78 GHz (at $a$ = 450 and 275 nm) and 1.58 GHz ($a$ = 600 nm), giving in average the value of 1.1 GHz. We attribute that peak to the propagating Rayleigh SAW ($k \neq 0$), i.e., the wavevector determined by the experimental setup (spot size of the pump and probe beams, N.A. of the lens or Py dot size). Roughly assuming that the excited Rayleigh SAW is at $k_0 = 1.55$ rad μm$^{-1}$ the frequency obtained from the dispersions in Figs. 4(a) and 4(b) fits the measured frequency 1.1 GHz well.

In general, the frequencies of all other modes decrease with the increasing separation between the Py dots. Interestingly, at the smallest $a$ ($a$ = 275 nm) they are at higher frequencies, covering the range from 10.40 to 15.90 GHz, and the range moves slightly to lower frequencies with increasing distance between the dots. At $a$ = 450 nm, the measured phonon peaks exist from 6.39 GHz to 14.34 GHz, and with further increase of $a$, the lowering of frequencies becomes rapid, reaching the range from 2.37 to 7.04 GHz at $a$ = 600 nm. The wavevectors of the excited acoustic waves are determined by the periodicity of the Py dots as also mentioned in Refs. [16], [17], [18], [21]. With the increase in interdot separation the wavevectors of excited acoustic waves decrease and hence the frequencies also decrease.

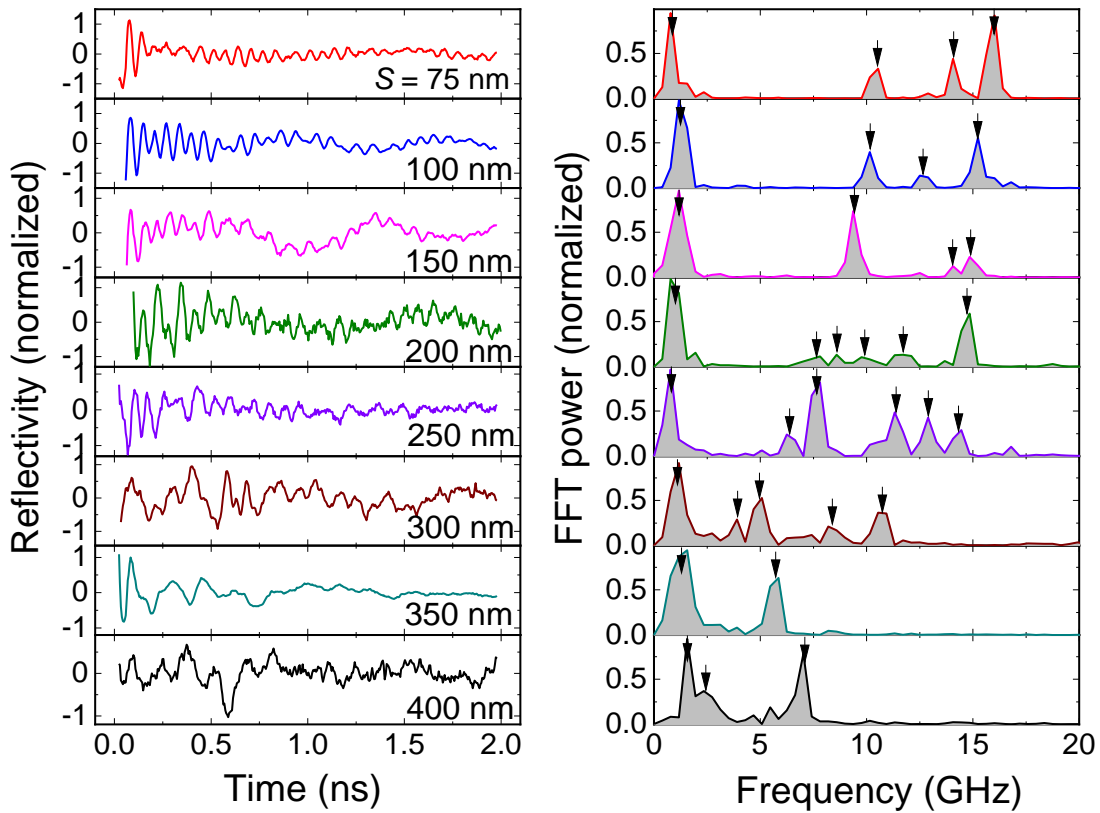

**Fig. 5.** Experimental spectra for the arrays of square Py dots with periods varying from 275 to 600 nm. The time domain spectra are shown in the left panel and their Fast Fourier Transform spectra are shown in the right panel. Arrows in the right panel indicate the resonance modes plotted in Fig. 6.

The measured dependence of the SAWs on the inverse of the lattice constant $q$ (and the lattice constant – upper scale) is compared with the frequency domain simulations. The results are shown in Fig. 6 with the frequencies extracted from Fig. 5 marked with white dots and as a color map of the simulation results, whose intensity is calculated according to Eq. (1). On the right-hand side of the Fig. 6, the *x*-component of the displacement of the recognized modes are visualized at the *xz* cross section of the unit cell, at $y = a/2$, i.e., between Py dots. We can identify the modes $R_i$ as Rayleigh and Sezawa waves, which are indicated by the phase change along the *z*-direction. In fact, $R_1$ is a 1st order Rayleigh mode with a single nodal line along the depth, $R_2$ is a 2nd order Rayleigh mode (or first Sezawa mode) also with a single nodal line inside the $SiO_2$ layer, and $R_3$ is a 3rd order Rayleigh mode (second Sezawa mode) with 2 nodal lines (one as in $R_2$ and another near the $SiO_2$/Si interface). The difference in Rayleigh mode and the first Sezawa mode is mainly in their opposite signs of the rotation of the displacements at the interface, i.e., the surface displacements are described by retrogressive ellipses for the Rayleigh mode, whereas they are progressive ellipses for the first Sezawa mode [8]. All of the modes have a nodal line parallel to the *z*-axis, indicating that they originate from the 2nd Brillion zone, i.e., for $k=2\pi/a$. This shows that the experimentally observed dependence *f(q)* for SAWs can be related to the dispersion relation of SAWs in the substrate, but contrary to Refs. [16], [17], they are not Rayleigh waves from different Brillouin zones, but rather different order Rayleigh type SAWs from the 2nd Brillouin zone. This is further supported by comparing the simulation results in Fig. 6 with the dispersion relations obtained from Fig. 4 in the limiting cases of the homogeneous $SiO_2$ layer on Si (at small *q*) and the $SiO_2$/Py bilayer on Si (at large *q*). Corresponding parts of these dispersion relations from Fig. 4 are superimposed on the color map in Fig. 6 with orange-dashed lines that follow the trend of the dispersion lines in the phononic crystals. To complete the analyses, we have also marked low-intensity modes from the edge of the third Brillouin zone ($R_1'$) and from the fourth Brillouin zone ($R_1''$) in Fig. 6, whose mode profiles are shown in the right part of the figure.

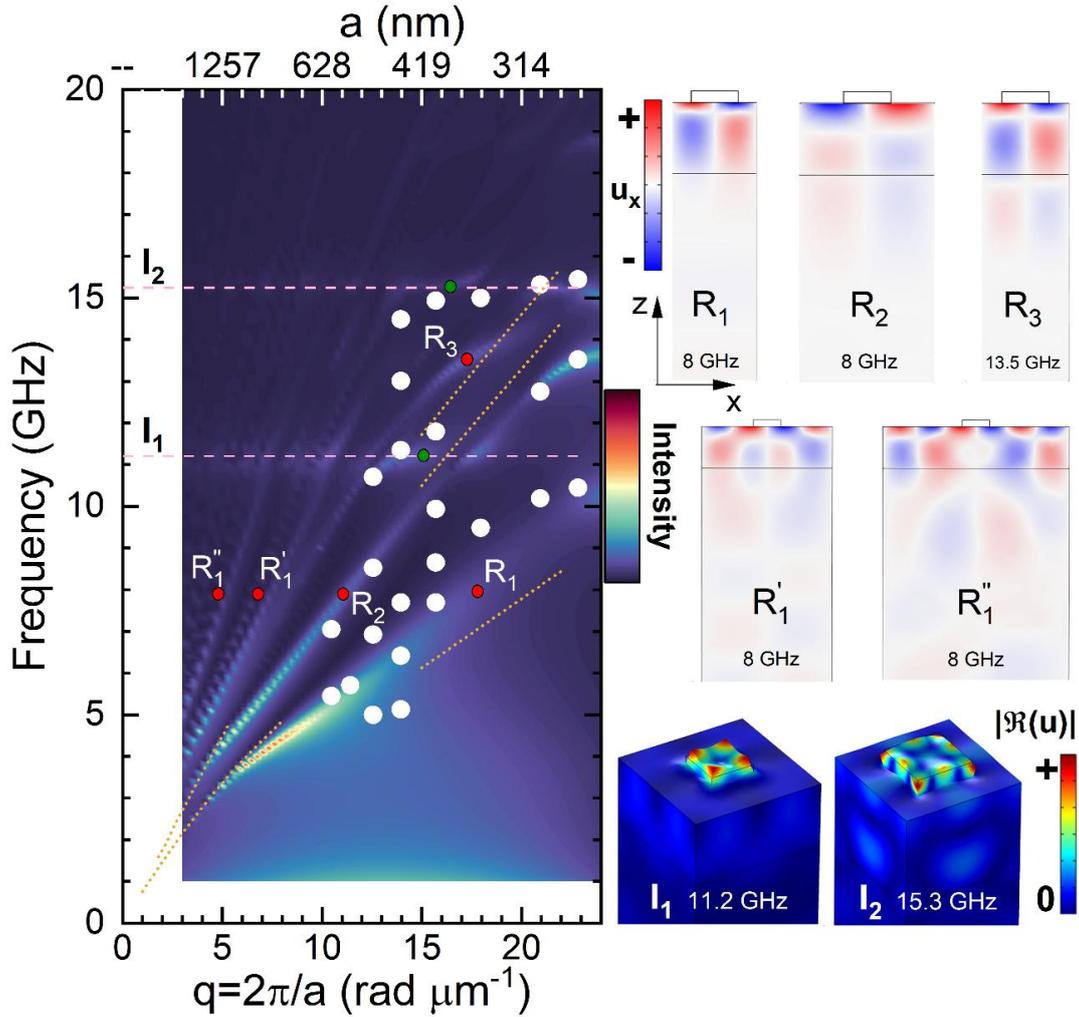

**Fig. 6.** The frequencies of the phonons extracted from the data presented in Fig. 5 (solid-white dots) and the modes of highest intensity from the simulations (color map) as a function of the inverse of the lattice constant (lower scale) or lattice constant (upper scale). The spectrum can be split into two family lines, one related to the SAWs in the substrate ($R_i$) and the other one to the confined modes of the Py dots ($I_i$). The dotted orange lines at low $q$ indicate the dispersion relations of $R_1$ and $R_2$ modes in a homogeneous $SiO_2$ layer on Si while at high $q$ they indicate the dispersion relations of $R_1$, $R_2$ and $R_3$ modes in a homogeneous Py/$SiO_2$ layer on Si (taken from Fig. 4). The mode profiles on the right show the $x$ component of the displacement for the $R_1$, $R_2$, and $R_3$ modes and the total displacement of localized modes $I_1$ and $I_2$ for Py dot, extracted from the simulations and marked on the color map with small red and green dots, respectively.

There are also other modes, labeled as $I_1$ and $I_2$, whose frequencies are almost independent on $q$ (except at the hybridizations with the SAWs as discussed below). These modes are confined to the Py dots, as confirmed by the profiles on the right part of Fig. 6. Careful inspection of these mode profiles reveals the oscillating wave component with decaying amplitude along the $z$ direction. It indicates the leakage of energy from the Py dots into the bulk modes of the substrate. An interesting feature is observed in the frequency range where the overlap between $R_i$ and I modes occurs, i.e. for small lattice constants, $a < 450$ nm ($q > 14$ rad $\mu m^{-1}$). In this frequency range the coupling between the phonons of the Py dots (I) and the SAWs propagating in the substrate ($R_i$) results in the hybridization between the modes and a change in their dispersion relations. Such effects are present in the simulation results as shown in Fig. 6. However, their experimental confirmation would require extensive additional experimental studies, in particular the fabrication of samples with small changes in the lattice constant.

To broaden the analysis of the phonon spectra in the considered systems, we performed eigenfrequency simulations, calculating the full phononic band structure along the irreducible path in the first Brillouin zone (along the path MΓXM) for two extreme values of the experimentally implemented lattice constants, $a = 275$ nm (Fig. 7) and $a = 600$ nm (Fig. 8). The spectra are denser than in Fig. 6, due to the selective excitation used in the frequency domain simulations in Fig. 6. The bands are colored according to the mode intensity $I$ calculated from Eq. (1), where red indicates high intensity and blue indicates low intensity. The continuum of the spectra, i.e., above the sound line, is marked in aquamarine color. At $k = k_0 = 1.55$ rad μm$^{-1}$, the estimated wavenumber selected in the measurement setup, only two bands are clearly separated from the band continuum. These are the Rayleigh and Love modes from the first Brillouin zone, with the profile of the Rayleigh mode (Rl) shown in Figs. 7b and 8b. The Sezawa and higher order Love modes are detached from the continuum at higher wavevectors and are therefore not detected in the first Brillouin zone in the measurements. Consequently, all measured modes, except the low frequency Rayleigh mode described above, overlap with the bulk continuum and originate in higher order Brillouin zones.

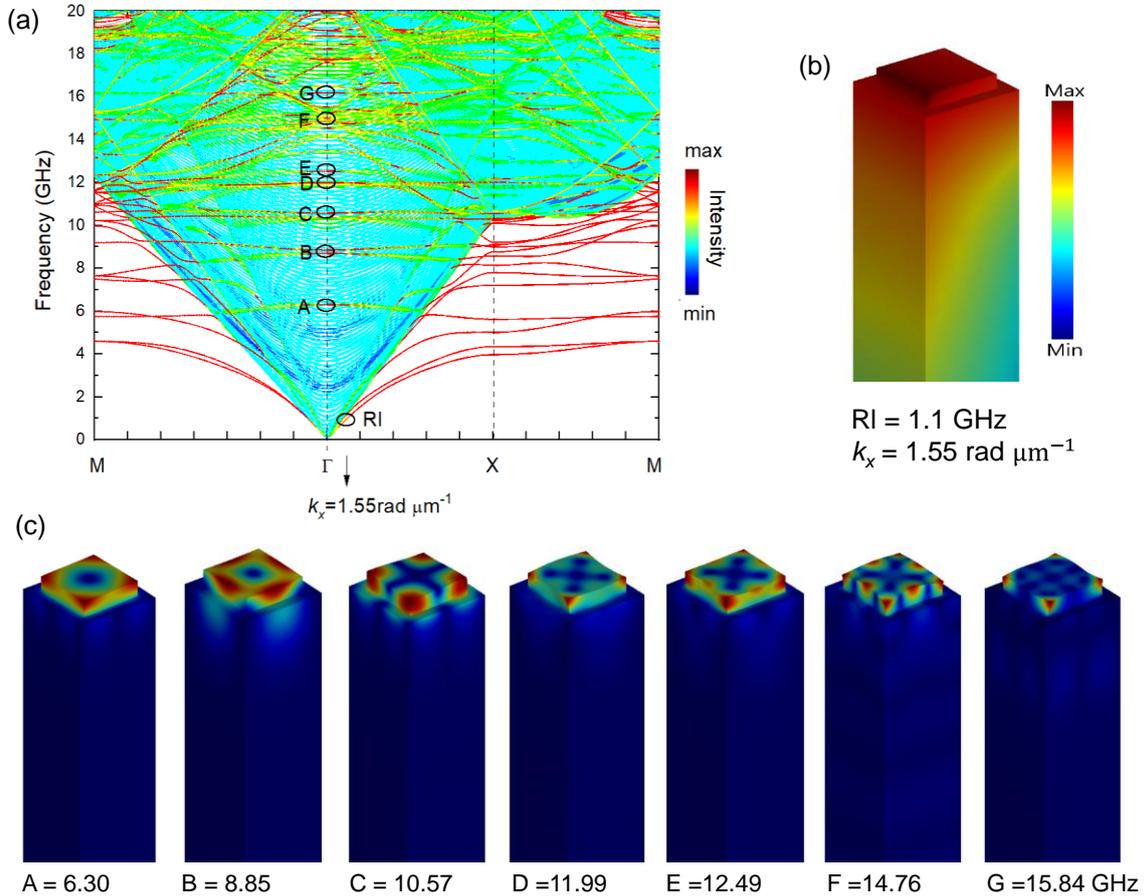

**Fig. 7.** (a) Phononic band structure in the phononic crystal with $a = 275$ nm ($q = 22.85$ rad μm$^{-1}$) calculated along the irreducible path (MΓXM) in the 1$^{st}$ Brillouin zone. The color indicates the mode intensity $I$ calculated according to Eq. (1) with red/yellow indicating the large values of the out-of-plane component of the displacement. The aquamarine color marks the continuum band of the bulk phonons. (b) Distribution of the total displacement of the Rayleigh SAW at $k_x = k_0 = 1.55$ rad μm$^{-1}$. (c) Distribution of the total component of the displacement in the unit cell at the center of the Brillouin zone ($k = 0$), for the modes with significant value of $I$, as marked in (a).

For $a = 275$ nm (Fig. 7(a)), all intense modes are in the range from 7.57 GHz (mode A) to 16.02 GHz (mode G). The distribution of the total displacement component of the modes labeled A–G, is shown in Fig. 7(c). These profiles are shown for $k = 0$ to simplify their interpretation, since the non-zero $k$ (e.g., $k_x = k_0$) introduces only the phase modulation along the direction of the $k$ vector. It is clear that all these modes have amplitudes concentrated in the Py dots, more precisely on the surfaces of the dots, with different quantization along the lateral directions. Since these modes overlap with the continuum band of bulk phonons, they leak energy into the substrate, as can be seen from the profiles in Fig. 7(c) and already seen in Fig. 6. This is the primary source of the attenuation of these Rayleigh and Sezawa waves [39].

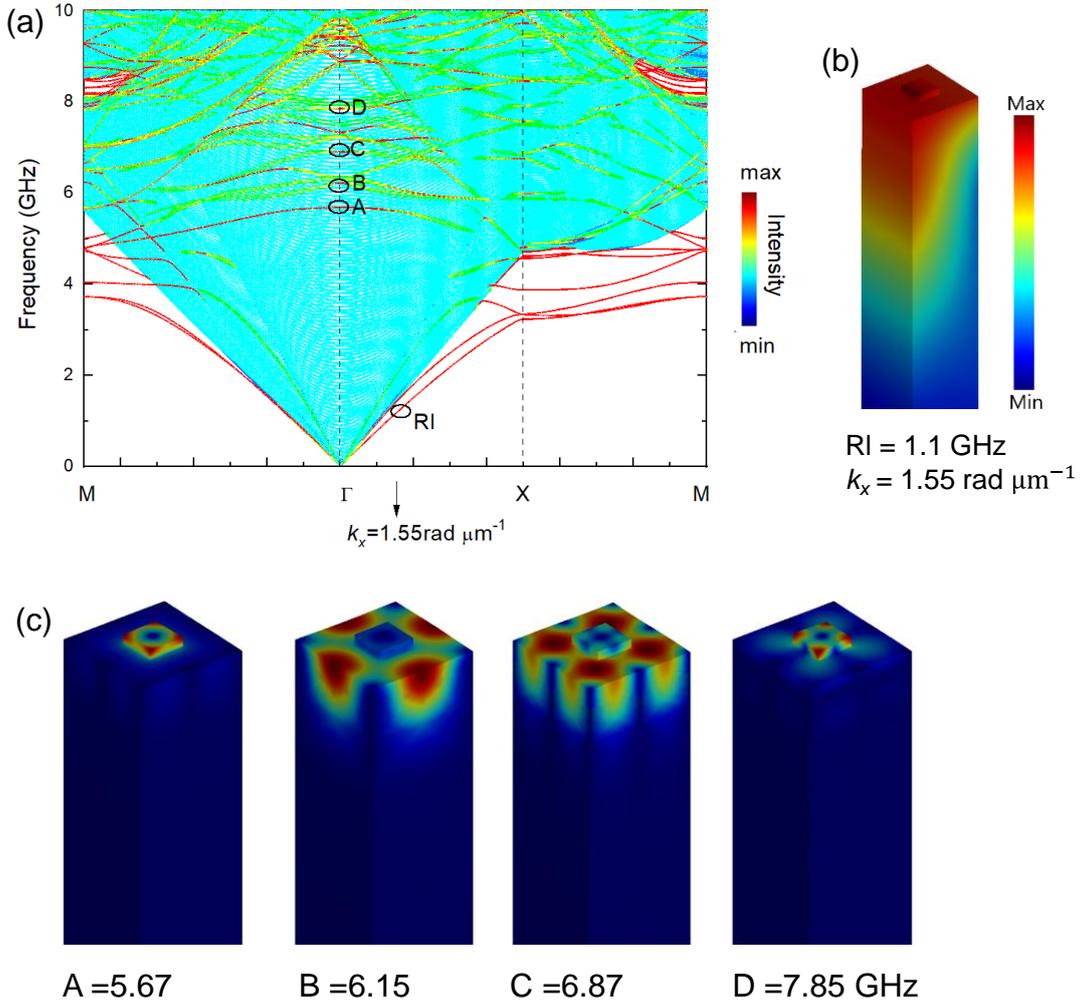

**Fig. 8.** (a) Phononic band structure in the crystal with $a = 600$ nm ($q = 10.41$ rad μm$^{-1}$) calculated along the irreducible path (MΓXM) in the 1$^{st}$ Brillouin zone. The color indicates the mode intensity $I$ calculated according to Eq. (1), with red/yellow indicating the large values of the out-of-plane component of the displacement. The aquamarine color marks the continuum band of the bulk phonons. (b) Distribution of the total displacement of the Rayleigh SAW at $k_x = k_0 = 1.55$ rad μm$^{-1}$. (c) Distribution of the total displacement in the unit cell at the center of the Brillouin zone ($k = 0$), for the modes with significant value of $I$, as marked in (a).

The band structure for the large lattice constant structure, $a = 600$ nm, shown in Fig. 8(a) is different from the previous one. The Brillouin zone is 2.2 times smaller, and the modes of high intensity are in a narrower frequency range from 6.45 GHz (mode A) to 7.62 GHz (mode D), in addition to the Rayleigh SAW [mode Rl, shown in Fig. 8(b)]. The bands associated with high intensity are more dispersive than the modes in Fig. 7, which indicates their propagative character. The total displacement profiles shown in Fig. 8(c) indicate that some of these modes are no longer concentrated in Py dots, but are distributed over the entire surface of the unit cell (modes B and C). Similar displacement distributions can be found in Fig. 6.

## 4. Conclusions

We study femtosecond pulsed laser-induced phonons in artificial crystals, i.e. phononic crystals, deposited on a multilayer dielectric substrate. The crystals consist of 50-nm-thick and 200-nm-wide square Py dots periodically arranged in a square lattice on a $Si/SiO_2$ substrate with varying interdot spacing, i.e., the lattice constant changing from 275 to 600 nm. A number of phononic modes are observed whose frequencies increase almost linearly with the inverse of the lattice constant, i.e., with $q = 2\pi/a$, for large values of $a$ ($a > 450$). For samples with smaller lattice constants, more complex spectra are observed with less significant dependence on $q$. The experimental results are interpreted with the aid of numerical simulations, in the frequency domain, and with the eigenfrequency solver. Three groups of measured modes are identified: Sezawa type of SAWs of different orders at frequencies from 5 to 15 GHz, the modes confined inside Py point at high frequencies (at 11.2 and 15.3 GHz), in addition to the fundamental Rayleigh mode at low frequency (1.1 GHz), whose frequency is almost independent of $a$. We show that the measured high frequency SAWs, with different quantization numbers over the thickness, originate from the second Brillouin zone, thus overlapping with the continuum of the substrate bulk modes. Furthermore, the simulation results indicate the possibility of hybridization between SAW and Py-dot confined modes. Thus, our study demonstrates a method to excite high-frequency Sezawa-type SAWs by ultrashort laser pulses and furthermore means to control the frequency of phononic modes by individual metallic nanoelements arranged in a periodic array by the inter-element spacing and the type of substrate. Since, ferromagnetic metal has been used to decorate the $Si/SiO_2$ substrate, the structure can also be useful for future exploitation of magnon-phonon couplings, further extending the usefulness of the investigated structures.

## Acknowledgments

The study has received from the National Science Centre of Poland under Grant No. UMO-2016/21/B/ST3/00452 and UMO-2020/37/B/ST3/03936. BR acknowledges NCN SONATA-16 project with Grant Number UMO-2020/39/D/ST3/02378. The authors sincerely thank Prof. YoshiChika Otani from RIKEN, Japan for preparing the samples for the study.

## References

[1] D. L. Arenberg, 'Ultrasonic Solid Delay Lines', *The Journal of the Acoustical Society of America*, vol. 20, no. 1, pp. 1–26, Jan. 1948, doi: 10.1121/1.1906343.
[2] C. Campbell, Ed., *Surface Acoustic Wave Devices and their Signal Processing Applications*. Academic Press, 1989. doi: 10.1016/B978-0-12-157345-4.50001-1.
[3] A. N. Darinskii and A. L. Shuvalov, 'Stoneley-type waves in anisotropic periodic superlattices', *Ultrasonics*, vol. 109, p. 106237, Jan. 2021, doi: 10.1016/j.ultras.2020.106237.
[4] A. Litvinenko, R. Khymyn, R. Ovcharov, and J. Åkerman, 'A 50-spin surface acoustic wave Ising machine', Nov. 12, 2023, *arXiv*: arXiv:2311.06830. doi: 10.48550/arXiv.2311.06830.

[5] L. Hackett *et al.*, 'Towards single-chip radiofrequency signal processing via acoustoelectric electron–phonon interactions', *Nat Commun*, vol. 12, no. 1, p. 2769, May 2021, doi: 10.1038/s41467-021-22935-1.

[6] T. Vasileiadis, J. Varghese, V. Babacic, J. Gomis-Bresco, D. Navarro Urrios, and B. Graczykowski, 'Progress and perspectives on phononic crystals', *Journal of Applied Physics*, vol. 129, no. 16, p. 160901, Apr. 2021, doi: 10.1063/5.0042337.

[7] Z. A. Alrowaili *et al.*, 'Heavy metals biosensor based on defective one-dimensional phononic crystals', *Ultrasonics*, vol. 130, p. 106928, Apr. 2023, doi: 10.1016/j.ultras.2023.106928.

[8] G. W. Farnell and E. L. Adler, *2 - Elastic Wave Propagation in Thin Layers*, vol. 9. ACADEMIC PRESS, INC., 1972. doi: 10.1016/B978-0-12-395670-5.50007-6.

[9] F. Hadj-Larbi and R. Serhane, 'Sezawa SAW devices: Review of numerical-experimental studies and recent applications', *Sensors and Actuators A: Physical*, vol. 292, pp. 169–197, Jun. 2019, doi: 10.1016/j.sna.2019.03.037.

[10] S. Büyükköse, B. Vratzov, D. Ataç, J. van der Veen, P. V. Santos, and W. G. van der Wiel, 'Ultrahigh-frequency surface acoustic wave transducers on ZnO/SiO2/Si using nanoimprint lithography', *Nanotechnology*, vol. 23, no. 31, p. 315303, Jul. 2012, doi: 10.1088/0957-4484/23/31/315303.

[11] H. Marom, J. Mullin, and M. Eizenberg, 'Size-dependent resistivity of nanometric copper wires', *Phys. Rev. B*, vol. 74, no. 4, p. 045411, Jul. 2006, doi: 10.1103/PhysRevB.74.045411.

[12] T. Sun *et al.*, 'Surface and grain-boundary scattering in nanometric Cu films', *Phys. Rev. B*, vol. 81, no. 15, p. 155454, Apr. 2010, doi: 10.1103/PhysRevB.81.155454.

[13] Y. C. Cho *et al.*, 'Copper Better than Silver: Electrical Resistivity of the Grain-Free Single-Crystal Copper Wire', *Crystal Growth & Design*, vol. 10, no. 6, pp. 2780–2784, Jun. 2010, doi: 10.1021/cg1003808.

[14] M.-H. Lu, L. Feng, and Y.-F. Chen, 'Phononic crystals and acoustic metamaterials', *Materials Today*, vol. 12, no. 12, pp. 34–42, Dec. 2009, doi: 10.1016/S1369-7021(09)70315-3.

[15] H. Yu *et al.*, 'Omnidirectional spin-wave nanograting coupler', *Nature Communications*, vol. 4, p. 2702, 2013, doi: 10.1038/ncomms3702.

[16] J.-F. Robillard, A. Devos, and I. Roch-Jeune, 'Time-resolved vibrations of two-dimensional hypersonic phononic crystals', *Phys. Rev. B*, vol. 76, no. 9, p. 092301, Sep. 2007, doi: 10.1103/PhysRevB.76.092301.

[17] J.-F. Robillard, A. Devos, I. Roch-Jeune, and P. A. Mante, 'Collective acoustic modes in various two-dimensional crystals by ultrafast acoustics: Theory and experiment', *Phys. Rev. B*, vol. 78, no. 6, p. 064302, Aug. 2008, doi: 10.1103/PhysRevB.78.064302.

[18] P. A. Mante, J. F. Robillard, and A. Devos, 'Complete thin film mechanical characterization using picosecond ultrasonics and nanostructured transducers: experimental demonstration on SiO2', *Applied Physics Letters*, vol. 93, no. 7, p. 071909, Aug. 2008, doi: 10.1063/1.2975171.

[19] *Computational Ocean Acoustics*. Accessed: May 08, 2024. [Online]. Available: https://link.springer.com/book/10.1007/978-1-4419-8678-8

[20] N. Favretto-Cristini *et al.*, 'Assessment of Risks Induced by Countermining Unexploded Large-Charge Historical Ordnance in a Shallow Water Environment—Part II: Modeling of Seismo-Acoustic Wave Propagation', *IEEE Journal of Oceanic Engineering*, vol. 47, no. 2, pp. 374–398, Apr. 2022, doi: 10.1109/JOE.2021.3111791.

[21] A. Comin *et al.*, 'Elastic and Magnetic Dynamics of Nanomagnet-Ordered Arrays Impulsively Excited by Subpicosecond Laser Pulses', *Phys. Rev. Lett.*, vol. 97, no. 21, p. 217201, Nov. 2006, doi: 10.1103/PhysRevLett.97.217201.

[22]   L. McKeehan and P. Cioffi, 'Magnetostriction in Permalloy', *Physical Review*, vol. 28, no. 2, Art. no. 2, 1926, doi: 10.1103/PhysRev.28.146.

[23]   E. Klokholm and J. A. Aboaf, 'The saturation magnetostriction of permalloy films', *Journal of Applied Physics*, vol. 52, pp. 2474–2476, 1981, doi: 10.1063/1.328971.

[24]   H. Pan *et al.*, 'Phononic and magnonic dispersions of surface waves on a permalloy/BARC nanostructured array.', *Nanoscale Research Letters*, vol. 8, no. 1, Art. no. 1, 2013, doi: 10.1186/1556-276X-8-115.

[25]   P. Graczyk, J. K\los, and M. Krawczyk, 'Broadband magnetoelastic coupling in magnonic-phononic crystals for high-frequency nanoscale spin-wave generation', *Phys. Rev. B*, vol. 95, no. 10, Art. no. 10, Mar. 2017, doi: 10.1103/PhysRevB.95.104425.

[26]   Z. Liu, F. Giesen, X. Zhu, R. D. Sydora, and M. R. Freeman, 'Spin Wave Dynamics and the Determination of Intrinsic Damping in Locally Excited Permalloy Thin Films', *Phys. Rev. Lett.*, vol. 98, no. 8, p. 087201, Feb. 2007, doi: 10.1103/PhysRevLett.98.087201.

[27]   K. Perzlmaier, G. Woltersdorf, and C. H. Back, 'Observation of the propagation and interference of spin waves in ferromagnetic thin films', *Phys. Rev. B*, vol. 77, no. 5, p. 054425, Feb. 2008, doi: 10.1103/PhysRevB.77.054425.

[28]   Z. Q. Qiu and S. D. Bader, 'Surface magneto-optic Kerr effect', *Review of Scientific Instruments*, vol. 71, no. 3, pp. 1243–1255, Mar. 2000, doi: 10.1063/1.1150496.

[29]   O. Matsuda, M. C. Larciprete, R. Li Voti, and O. B. Wright, 'Fundamentals of picosecond laser ultrasonics', *Ultrasonics*, vol. 56, pp. 3–20, 2015, doi: 10.1016/j.ultras.2014.06.005.

[30]   O. Matsuda, T. Pezeril, I. Chaban, K. Fujita, and V. Gusev, 'Time-domain Brillouin scattering assisted by diffraction gratings', *Phys. Rev. B*, vol. 97, no. 6, p. 064301, Feb. 2018, doi: 10.1103/PhysRevB.97.064301.

[31]   A. A. Maznev, J. Hohlfeld, and J. Güdde, 'Surface thermal expansion of metal under femtosecond laser irradiation', *Journal of Applied Physics*, vol. 82, no. 10, pp. 5082–5085, 1997, doi: 10.1063/1.366382.

[32]   B. Rana, S. Pal, S. Barman, Y. Fukuma, Y. Otani, and A. Barman, 'All-Optical Excitation and Detection of Picosecond Dynamics of Ordered Arrays of Nanomagnets with Varying Areal Density', *Appl. Phys. Express*, vol. 4, no. 11, p. 113003, Oct. 2011, doi: 10.1143/APEX.4.113003.

[33]   B. Rana *et al.*, 'Detection of Picosecond Magnetization Dynamics of 50 nm Magnetic Dots down to the Single Dot Regime', *ACS Nano*, vol. 5, no. 12, pp. 9559–9565, Dec. 2011, doi: 10.1021/nn202791g.

[34]   B. Graczykowski, S. Mielcarek, A. Trzaskowska, J. Sarkar, P. Hakonen, and B. Mroz, 'Tuning of a hypersonic surface phononic band gap using a nanoscale two-dimensional lattice of pillars', *Phys. Rev. B*, vol. 86, no. 8, p. 085426, Aug. 2012, doi: 10.1103/PhysRevB.86.085426.

[35]   B. Graczykowski, F. Alzina, J. Gomis-Bresco, and C. M. Sotomayor Torres, 'Finite element analysis of true and pseudo surface acoustic waves in one-dimensional phononic crystals', *Journal of Applied Physics*, vol. 119, no. 2, p. 025308, Jan. 2016, doi: 10.1063/1.4939825.

[36]   P. Graczyk and B. Mroz, 'Simulations of acoustic waves bandgaps in a surface of silicon with a periodic hole structure in a thin nickel film', *AIP Advances*, no. 7, p. 077138, 2014, doi: 10.1063/1.4892076.

[37]   A. Trzaskowska, S. Mielcarek, M. Wiesner, F. Lombardi, and B. Mroz, 'Dispersion of the surface phonons in semiconductor/topological insulator Si/Bi2Te3 heterostructure studied by high resolution Brillouin spectroscopy', *Ultrasonics*, vol. 117, p. 106526, Dec. 2021, doi: 10.1016/j.ultras.2021.106526.

[38]   B. Graczykowski, A. Gueddida, B. Djafari-Rouhani, H.-J. Butt, and G. Fytas, 'Brillouin light scattering under one-dimensional confinement: Symmetry and interference self-canceling', *Phys. Rev. B*, vol. 99, no. 16, p. 165431, Apr. 2019, doi: 10.1103/PhysRevB.99.165431.

[39]   D. Li and D. G. Cahill, 'Attenuation of 7 GHz surface acoustic waves on silicon', *Phys. Rev. B*, vol. 94, no. 10, p. 104306, Sep. 2016, doi: 10.1103/PhysRevB.94.104306.